# Hypercomputing the Mandelbrot Set?

Petrus H. Potgieter[*]

September 3, 2018

**Abstract:** The Mandelbrot set is an extremely well-known mathematical object that can be described in a quite simple way but has very interesting and non-trivial properties. This paper surveys some results that are known concerning the (non-)computability of the set. It considers two models of decidability over the reals (which are treated much more thoroughly and technically in [1], [2], [3] and [4] among others), two over the computable reals (the Russian school and hypercomputation) and a model over the rationals.

## 1 Introduction

In theoretical computer science it is no surprise that the halting problem for Turing machines is the favourite target for solution by non-conventional models of computation. However, the decidability of sets of reals and the computability of functions in ordinary real analysis is a topic of great interest to the broader mathematical community and a potential area of application for so-called hypercomputation. This paper surveys some of the most important results and gives an extremely simple example of the application of accelerated Turing machines to the question of decidability of the Mandelbrot set. This very amusing problem was raised by Roger Penrose [5] and—like many good questions—implies considerable work on the definitions, in this case what exactly is meant by *decidable* for a subset of the plane.

In 1979 Benoît Mandelbrot used a computer to plot[1] a beautiful approximation of the subset

$$M = \{c \in \mathbb{C} \mid \text{for all } n \geq 1,\ |f_c^n(0)| \leq 2\} \qquad \text{where } f_c(x) = x^2 + c \quad (1)$$

of the complex plane $\mathbb{C}$ (where $f^n$ denotes $n$-th iteration of $f$). This set was originally described by Pierre Fatou in 1905 but after the appearance of a colourful plot of the set in Mandelbrot's book [6] and—especially—on the cover of Scientific American and in an accompanying column [7] in August 1985, the Mandelbrot set has become one of the greatest celebrities of popular mathematics.

---
[*]Department of Decision Sciences, University of South Africa (Pretoria), PO Box 392, UNISA, 0003, Republic of South Africa, potgiph@unisa.ac.za.
[1]Mandelbrot actually plotted a *mirror image* of $M$.

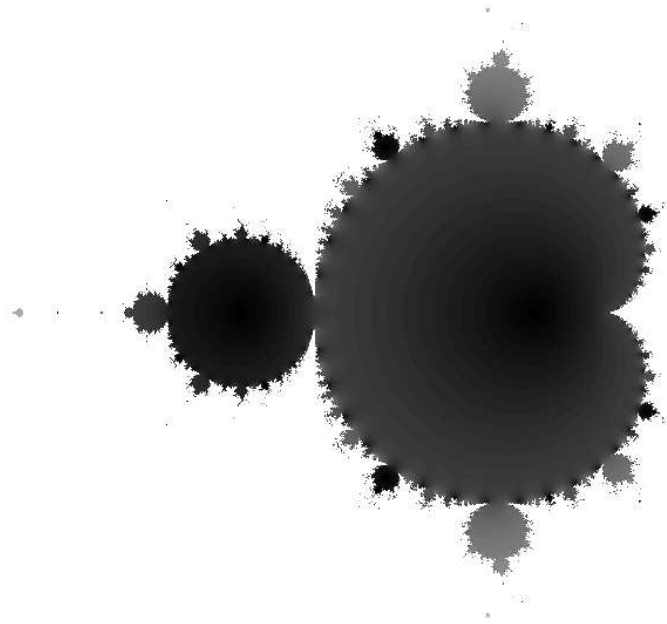

Figure 1: Membership candidates for the Mandelbrot set

Plots can easily be generated using a plethora of software freely available, including normal LaTeX code[2]. The relatively concise but non-optimised Octave[3] code snippet

```
n=1000;                           # For an nxn grid
m=50;                             # Number of iterations
c=meshgrid(linspace(-2,2,n))\     # Set up grid
  +i*meshgrid(linspace(2,-2,n))';
x=zeros(n,n);                     # Initial value on grid
for i=1:m
   x=x.^2+c;                      # Iterate the mapping
endfor
imagesc(min(abs(x),2.1))          # Plot monochrome, absolute
                                  # value of 2.1 is escape
```

suffices, for example, to plot the Mandelbrot image in Figure 1.

The image in Figure 1 shows an expanse of background (white) points which were shown during the execution of the code to have left the closed disk of

---

[2]http://www.thole.org/manfred/apfel/apfel.tex [accessed 2005-12-30] by Manfred Thole. The Mandelbrot set is approximated by the white area at the centre of this plot generated by a correctly compiled `apfel.tex`.

[3]Octave is a free and open-source high-level language for numerical computation with implementations on many platforms and largely compatible with MATLAB®. See http://www.octave.org/.

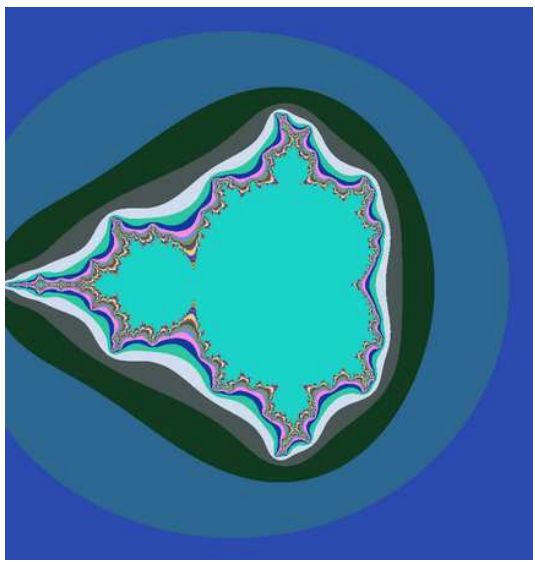

Figure 2: The Mandelbrot plot as we too often see it radius 2—therefore not to belong to the Mandelbrot set—and grey and black points which are *candidates* for membership of $M$. In this plot the points $c$ have a lighter colour when they have approached relatively closely to the boundary of the disk after exactly 50 iterations of the map $f_c(0)$. This shading scheme emphasises that the usual routines for generating Mandelbrot plots, like this one, clearly can identify only elements of the complement of $M$ and not of the set itself. Scientists are well-aware this fact (that the complement of $M$ is only known to be recursively enumerable) but it is usually not overemphasised in the more popular writing. Incidentally, $M$ is compact and simply connected [8] which one cannot, and should not expect to, see in Figure 1. Figure 2 shows a plot of the Mandelbrot set candidates as one more usually sees it.

The central part is, of course, the same as the set of grey or black points in Figure 1—those points who have not (yet) escaped the disk after $m$ iterations in the Octave code fragment

```
n=1000;                       # For an nxn grid
m=60;                         # Number of iterations
c=meshgrid(linspace(-2,2,n))\ # Set up grid
  +i*meshgrid(linspace(2,-2,n))';
x=zeros(n,n);y=ones(n,n);     # Initial values on grid
                              # y counts number of iterations
                              # without escape from the disk
for i=1:m
   x=x.^2+c;                  # Usual iteration
```

```
        y=y.+(.5+sign(3-abs(x))./2);# Add one if still in radius 3
        x=x.*(min(abs(x),3)\          # Scale back points far away to
          ./(abs(x)+!abs(x)));        # speed up and avoid overflow
     endfor
     imshow(imagesc(y),rand(m+1,3)) # Plot y with random colours
```

which now keeps track of when a point first leaves the disk (if at all) using the matrix $y$. The union of the outer coloured bands represents $\mathbb{R} \setminus M$ on the computational grid, or what we could discover of it in the number of iterations executed. Although pretty, these bands show the stages of construction of $M$'s complement and disguise the fact that we know about $M$ not much more than that it is somewhere in the central coloured section! Incidentally, John Ewing and Glenn Schober have made [9] laborious numerical estimates of the area[4] of the Mandelbrot set using two different plotting methods and arrived at answers of respectively 1.52 and 1.72. This is for a set which everybody believes that they have seen!

In a very readable article [11] Lenore Blum has described the gulf between numerical analysis and computer science. Blum calls the classical theory of (Turing) computability

> fundamentally inadequate for providing such a foundation for modern scientific computation, in which most algorithms–with origins in Newton, Euler, Gauss, et al.–are real number algorithms as a partial justification of the Blum-Shub-Smale model of computation over the real numbers (mentioned in 3.1, below). The present author would be inclined to the opinion that 'modern scientific computation' is rather inadequately founded in the classical theory of computability and that the logical and foundational problems which arise in this regard (some of which are illustrated in this paper) could be much deeper and connected to our model of the real numbers and their representation; and to the finite/infinite duality which is drilled into mathematical recruits in a dingy room in the Hilbert Grand Hotel.

## 2    $M$ in the Recursive Realm

First, let us consider only computable points in the Mandelbrot set $M$. By this we mean points that can be finitely specified and communicated in a consistent way, which one can take to mean *represented by a program for computing approximating rationals*. This is an intuitive idea used in all versions of computable analysis and which can be readily grasped. Fix a universal Turing machine $U$ as well as a (recursive, surjective) encoding $\phi : \mathbb{N} \to \{0, 1\} \times \mathbb{N} \times \mathbb{N}$ with

$$\phi : n \mapsto (\phi_1(n), \phi_2(n), \phi_3(n))$$

so that every rational number is of the form

$$(-1)^{\phi_1(n)} \frac{\phi_2(n) - 1}{\phi_3(n)}$$

---

[4] A tighter estimate has recently been computed by Yuval Fisher and Jay Hill [10].

for some $n$. If $x \in \mathbb{R}$ then we say that $x$ belongs to the set $\mathbb{R}_c$ of computable reals if there exists a program[5] $i_x$ for $U$ so that, if $f_{i_x}$ is the function computed by $i_x$ then

$$\left| x - (-1)^{\phi_1(f_{i_x}(m))} \frac{\phi_2(f_{i_x}(m)) - 1}{\phi_3(f_{i_x}(m))} \right| < 2^{-m}$$

for every $m \in \mathbb{N}$. Penrose's question in this section should be:

> Given a program-description of a computable complex number, can we algorithmically determine whether it belongs to $M$, or not?

In other words, does there exist a (partial) function $G :\subseteq \mathbb{N} \times \mathbb{N} \to \mathbb{N}$, computable in some sense, such that whenever $i_x$ and $i_y$ are programs for $x, y \in \mathbb{R}_c$ respectively then

$$G(i_x, i_y) = \chi_M(x + iy). \qquad (2)$$

This simply means that $G$ can be used to determine membership of $M$ for the recursive points in the plane. Identify $\mathbb{C}$ with $\mathbb{R}^2$ in the usual way and consider $M_c = M \cap \mathbb{R}_c^2$, the computable points in the plane that belong to the Mandelbrot set. The unsuspecting reader of popular scientific literature could reasonably assume that given a full description of a point $(x, y)$, by a program pair $(i_x, i_y)$ one supposes there exists a procedure for deciding membership of $M_c$. The existence of such a $G$ has been implicitly suggested to the general public for two decades by the pretty pictures we have had our computers draw, but it turns out that we do not have such a procedure in classical computability theory at all.

## 2.1 Markov computability

In the Russian school of constructive mathematics, pioneered by Andrei Markov, a function $f : \mathbb{R}_c \to \mathbb{R}_c$ is considered to be computable (or, constructive) if there exists a Turing machine computable $G : \mathbb{N} \to \mathbb{N}$ such that whenever $i_x$ is a $U$-program for $x$ then $G(i_x)$ is a $U$-program for $f(x)$—and $f(x)$ is defined if and only if $G(i_x)$ is. A set is computable or decidable in this setting when its characteristic function is computable. However, every computable function is continuous [12] and therefore the only computable sets will be closed (and open). This is a rather ironic development as, for example, the computable interval $[0, 1] \cap \mathbb{R}_c$ will not be a computable set in this sense, being closed but not open in $\mathbb{R}_c$. The same is true of the unit circle and the disk in the plane and in both cases the difficulty springs from the fact that there is not general procedure that, given $i_x$ and $i_y$, can decide whether $x = y$ or not. The Mandelbrot set is, however, closed in $\mathbb{R}$ (and hence closed in $\mathbb{R}_c$), so could it be computable in this sense? No, since $-2 \in M_c$ and $-2$ is a cluster point in $\mathbb{R}_c$ of $\mathbb{R}_c \setminus M_c$ and hence $M_c$ is not open in $\mathbb{R}_c$. Consequently its characteristic function cannot be Markov computable[6].

---

[5]We shall assume that the *program* and *input* for $U$ (actually both simply inputs) are natural numbers.

[6]Since the unit circle is also not computable here, this should not come as a great surprise.

Although apparently intuitive, the notion of computable set used in this subsection is clearly very, very bad from the point of view of real analysis. Nevertheless it corresponds in some sense exactly to what a programmer would regard as computable: given a procedure (subroutine) for an $x$, having a program that outputs 1 in finite time if $x \in M_c$ and zero otherwise. The situation here is in contrast to that of subsets of the natural numbers, for which the notion of decidability is very natural and well-established (*pace* the entire field of 'super-Turing' hypercomputation).

## 2.2 Zeno Machine Computability

Consider what may be called a *Zeno machine* (ZM) [13] or Accelerated Turing Machine [14, 15]. With this kind of speed-up of the computing device, one can solve the halting problem for Turing machines in finite time. Many hypercomputational schemes tend to be proposals for somehow accomplishing infinitely many computational steps in a finite time (see also [16], for example). Without loss of generality, we shall assume a ZM to be identical to a Turing machine with one input tape, one output tape and a storage tape *except* that the ZM takes $\frac{1}{2}$ hour to execute the first transition, $\frac{1}{4}$ hour for the second, $\frac{1}{8}$ hour for the third etc. After one hour the ZM will have finished its operation and one will perhaps find the answer to some tantalising question on the output tape. On a putative ZM one could implement an Octave interpreter that would execute the code

```
n=1000;                        # For an nxn grid
c=meshgrid(linspace(-2,2,n))\  # Set up grid
 +i*meshgrid(linspace(2,-2,n))';
x=zeros(n,n);                  # Initial value on grid
do
   x=x.^2+c;                   # Usual iteration
   x=x.*(min(abs(x),3)\        # Scale back points far away to
    ./(abs(x)+!abs(x)));       # speed up and avoid overflow
                               # and infinite values
until (1==0)                   # Repeat a lot
imagesc(min(abs(x),2.1))       # Plot x, 2.1 counts as escape
```

in finite time. This would provide an *exact* plot of the Mandelbrot set on the grid points! With a small modification (adding a procedure/subroutine for approximating the computable real) a ZM can decide membership of $M_c$ for any computable real. For example, the Octave code immediately above can be rewritten for an (ordinary, Turing computable) Octave function $c(i)$ (instead of a matrix $c$) where $c(i)$ gives a rational approximation of $c$ to within $2^{-i}$ and a scalar $x$, initially zero. In the $i$-th iteration of the loop we then recompute[7] $x$ using $c(i)$. A similar calculation could be done relative to an oracle for the halting problem. The ZM as used here does not *apparently* present any of the problems with respect to defining the terminal configuration of the devices described in [13] since the matrix $x$ is always bounded. This stability is however illusory: in executing the code described here one needs to continually reset $x$ back to zero and therefore the variable $x$ will, for every $x \in \mathbb{R}_c \setminus M_c$, have values alternatively 0 and with absolute value 2.1 arbitrarily close to the end of execution time[8].

---
[7]This would require recomputing the previous iterates, of course, but there are only finitely many to do each time.

## 2.3 A Rational Refuge?

Is there some relief from these problems if we restrict our attention to the points with rational coordinates only? Consider again the (recursive, surjective) encoding $\phi : \mathbb{N} \to \{0, 1\} \times \mathbb{N}^2$ of the rational numbers used earlier. A set $A \subseteq \mathbb{Q}$ can be defined as computable whenever a Turing machine computable function $F : \mathbb{N} \to \mathbb{N}$ exists such that

$$F|_{\phi^{-1}(A)} \equiv \chi_A \circ \phi.$$

In this sense, now, the rational points on the unit circle do constitute a computable set since the condition $x^2 + y^2 = 1$ can be checked by a Turing machine for rational $x$ and $y$. Is $M \cap \mathbb{Q}$ computable in *this* sense?

This kind of computability over the rationals is very different from Markov-computability over the computable reals. Consider for example the function $f : \mathbb{Q} \to \{0, 1\}$ such that $f(q) = 1$ if the reduced improper fraction representation of $q$ has an even denominator and $f(q) = 0$ otherwise. This function is not continuous on $\mathbb{Q}$ and therefore not the restriction of a Markov-computable function to the rationals. It is therefore conceivable[9] that a Turing machine could compute the characteristic function of $M \cap \mathbb{Q}$ with respect to the representation $\phi$ of the rationals, $F|_{\phi^{-1}(M \cap \mathbb{Q})}$.

The rational points are perhaps a bad basis for developing a general theory of computability of subsets of $\mathbb{R}^n$ however. It could say nothing much about the curve $x^3 + y^3 = 1$. In fact, decidability with respect to the rationals suffers from a general failure to take the boundary into account (as in the example below). Nevertheless, for connected and compact sets with non-empty interior, computability in this sense seems a relatively natural and quite desirable property. It would, for instance, allow one to plot the set with a computer using test points on a rational grid. Consider also that the set $\{(x, y) \in \mathbb{Q}^2 \mid y \geq e^x\}$ is computable with respect to the representation $\phi$: if $e_1(q, m)$ is a computable function approximation of $e^q$ from below and $e_2(q, m)$ from above such that $\lim_{m \to \infty} e_i(q, m) = e^q$ then an enumeration of the values $e_1(q, m)$ and $e_2(q, m)$ for $m = 1, 2, \ldots$ will after finitely many steps reveal whether any given rational lies below $e^q$ or above it (since $e^q$ is irrational for all rational $q \neq 0$ and the case $q = 0$ can be checked separately).

---

[8]Thomson's Lamp showing the way...

[9]Although it strikes the present author as unlikely that $M$ will be computable in this sense, a proof is called for.

# 3  $M$ in Real Space

Let us return to the standard real numbers and consider $M$ as a subset of the standard plane. In pursuing an answer to Penrose's question, the formulation of an appropriate notion of decidability of subsets is again required. Most of the work in this regard has its roots in the Polish school of computable analysis, starting with Andrzej Gregorczyk and Daniel Lacombe. The Blum-Shub-Smale system on the other hand, has a rather algebraic flavour.

## 3.1  The Blum-Smale Result

Blum, Mike Shub and Steven Smale (BSS) have introduced [2, 11] a model of computation over arbitrary commutative rings which is based on machines that operate using the elements of the ring $R$ in lieu of a finite alphabet. There exist universal machines in this model. In the case where the ring is $\mathbb{Z}_2$ the classical computability theory is recovered. BSS-computable functions over $R$ are functions computed by such a machine and they call a set computable or decidable whenever its characteristic function is BSS-computable. Blum and Smale have shown that the Mandelbrot set is NOT computable in this framework [17] which is at least a partial answer to Penrose's question. However, Vasco Brattka has shown [3] that in the BSS scheme over the field of standard real numbers the set $\{(x,y) \in \mathbb{R}^2 \mid y \geq e^x\}$ is not computable either. Brattka's result reflects unfavourably on the claim that BSS computability provides a natural notion of decidable set—at least, for real analysis.

## 3.2  Computable Real Analysis

Computable real analysis in the Polish school, as developed in [18] and elsewhere, is based on the definition of a function as computable whenever it maps every computable sequence of points (in $\mathbb{R}^n$) to a computable sequence of points and has a recursive modulus of continuity (defined on $\{1, \frac{1}{2}, \frac{1}{3}, \frac{1}{4}, \ldots\}$) on every compact subset. The appropriate definition [3] in this context is for a set $A$ to be *computable* whenever its distance function $d_A : \mathbb{R}^n \to \mathbb{R}$ is a computable function in this sense. Computable (in the classical sense) subsets of $\mathbb{N}$, viewed as subsets of $\mathbb{R}$, remain computable in this sense [1] and the notion is therefore a true generalisation of the notion of classical Turing computability. Peter Hertling has recently shown [1] that if the Mandelbrot set $M$ is locally connected[10], then its distance function has to be computable and hence Penrose's question would be answered in the affirmative.

As described in [4], in this notion of computability the closed decidable subsets of $\mathbb{R}^n$ are *exactly* those sets which can in principle be plotted with arbitrary accuracy on a computer screen[11]. It is therefore at this point not yet known whether the Mandelbrot set can be accurately drawn by a computer!

---

[10] Actually, Hertling proved a stronger result: that the hyperbolic conjecture (which would be implied by local connectedness) would be sufficient to prove imply computability of the distance function.

## 4 Conclusion

The table below summarises the results and the two open problems mentioned in this survey.

|                                      | Circle | $y \geq e^x$ | M'brot |
|--------------------------------------|:------:|:------------:|:------:|
| Markov-computability over $\mathbb{R}_c$ | ×     | ×            | ×      |
| Blum-Shub-Smale over $\mathbb{R}$    | ✓      | ×            | ×      |
| Turing-computability over $\mathbb{Q}$ | ✓    | ✓            | ?      |
| Computable analysis (Brattka e.a.)    | ✓     | ✓            | ?      |
| Zeno-computability over $\mathbb{R}_c$ | ✓    | ✓            | ✓      |

Although the classical computability of the rational Mandelbrot set $M \cap \mathbb{Q}^2$ is an obvious question which one would cautiously expect to be answered in the negative, the author is not aware of a current result implying this. Among the models of decidability of sets in $\mathbb{R}^n$, the approach 3.2 studied by Brattka, Weihrauch e.a. appears the most reasonable and a demonstration of the computability of $M$ in this setting would be extremely interesting both in itself as well providing strong support for the intuitiveness of their approach. In the model over $\mathbb{Q}$ the Mandelbrot set is of course decidable with respect to an oracle for the halting problem but it seems unlikely the converse is true, so if $M \cap \mathbb{Q}^2$ is not Turing-decidable then it could be interesting[12] to study problems that can be solved relative to an oracle for $M \cap \mathbb{Q}^2$ (or an $M$-oracle in any model in which $M$ is not decidable). Until such time as at least one of the question marks in the table have been decided, a Zeno machine (or any hypercomputing model capable of solving the halting problem for Turing machines) remains— alas!—the best way of imagining that we can actually decide membership of the Mandelbrot set.

---

[11] Using an approach which switches a pixel *on* when the set is *close* to the centre of the pixel. This approach requires some obvious assumptions about the scale.

[12] This question was also raised by Klaus Meer and Martin Ziegler in slides for a talk, http://www.upb.de/cs/ag-madh/WWW/ziegler/LUEBECK2.pdf [accessed 2006-01-04].